\documentstyle[12pt]{article}
\normalbaselines
\def\zid{1\kern-0.36em\llap~1}

\newcommand{\beq}{\begin{equation}}
\newcommand{\ber}{\begin{eqnarray}}
\newcommand{\eeq}{\end{equation}}
\newcommand{\eer}{\end{eqnarray}}

\begin{document}

\begin{titlepage}
\vbox {\vspace{1cm}} %Leaves space at top of 1st page.
\rightline{[SUNY BING 9/26/04] } \rightline{ hep-ph/0410077}
\vspace{3cm}
\begin{center}
{\bf TESTS FOR THE STATISTICS OF PAIR-PRODUCED NEW
PARTICLES}\footnote{For Proceedings of ``Vth Rencontres du Vietnam
2004" .}\\ \vspace{2mm} Charles A.
Nelson\footnote{Electronic address: cnelson @ binghamton.edu  } \\
{\it Department of Physics, State University
of New York at Binghamton\\ Binghamton, N.Y. 13902}\\[2mm]
\end{center}

\vspace{2mm}

\begin{abstract}
Due to selection rules, new particles are sometimes
discovered/predicted to be produced in pairs.  In the current
search for SUSY particles this will occur if R-parity is
conserved.  In local relativistic field theory, there can be
identical particles which are neither bosons nor fermions which
are associated with higher-dimensional representations of the
permutation group.  Such particles will generally be pair-produced
and so empirical tests are required to exclude them.  A
parameter-free statistical model is used to study the unusual
multiplicity signatures in coherent paraboson production versus
the case of ordinary bosons.
\end{abstract}

\end{titlepage}

\section{Comparison of Selection Rules in SUSY (from R-Parity) and
Parastatistics}

The general motivation is the question ``Where are the fundamental
particles corresponding to the other representations of the
permutation group?" So far, fermions go in totally anti-symmetric
representations and bosons go in totally symmetric
representations. But there are mixed representations. Particles
associated with such representations obey parastatistics \cite{1}
. The order of the paraparticles is denoted by ``$p$" such that
``$p=1$" corresponds to the ordinary bosons and fermions. In
general, ``$p$" is the maximum number of parafermions (parabosons)
that can occur in a totaly symmetric (anti-symmetric) state.

\subsection{Selection Rules}

From R-parity, all supersymmetric particles are assigned a new
conserved quantum number
\begin{equation}
R=(-)^{3B+L+2J}
\end{equation}
so $R = \pm 1 $ respectively for particles and superparticles.
Consequently, superparticles are pair-produced and the lightest
superpartner (LSP) must be stable.

From locality, the selection rules \cite{1,2} for the production
of paraparticles are similar: Transitions with only one external
$p>1$ particle are forbidden, so the lightest such particle is
stable. The number of parafermions is conserved modulo 2. For a
$p$-even family, the number of external lines is even, so an even
number of $p>1$ must be produced. \ For a $p$-odd family, the
number of external lines cannot be an odd number less than $p$.
For instance, if $p=5$ , then there cannot be 3 paraparticles
produced. However, in the case of $p=3$ there is the important
exception that while a single paraparticle cannot be produced, the
production of 3 such particles is allowed.

\subsection{Cross-Section Tests}

In the case of SUSY, once the masses are known, the cross-sections
are fixed. \ For instance, sleptons and electroweak gauginos
interact with
strength $\alpha .$ \ Similarly, squarks and gluinos interact with strength $%
\alpha _{QCD}$ . \ In leading order in the coupling, the
corresponding cross-section in the pair-production of two
analogous order $p$ paraparticles is a factor of $p$ larger. \ In
comparison with SUSY, such a cross-section test can be used to
exclude the production of paraparticles. \ However, if a
cross-section were found to be a factor of $p$ greater than
expected, this would be the same prediction as in the case of the
production of $p$-fold degenerate fermions/bosons. In particular,
there is the issue ``How to distinguish $p=2$ parabosons from
two-fold degenerate ordinary bosons?'' In the case when SUSY is
not involved in the pair-production processes, the magnitude of
the cross-sections might not be known apriori, but this same
question arises.

\newpage

\section{Coherent Production of Parabosons of Order 2}

The multiplicity signatures for $p=2$ parabosons and ordinary
bosons are indeed different. To show this, we use a parameter-free
statistical model \cite{3} to study multiplicity signatures for
coherent production of charged-pairs of parabosons of order $p=2$
in comparison with those arising in the case of ordinary bosons,
$p=1$. This model gives 3D plots of the pair probability
${P_{m}}(q)$$\equiv$ `` the probability of $m$ paraboson
charged-pairs + $q$ positive parabosons" versus $<n>$ and $<n^2>$.
As shown in the figure from Ref. [3], the $p=1$ curve is found to
lie on the relatively narrow 2D $p=2$ surface. Such signatures
distinguish between $p=2$ parabosons and two-fold degenerate
ordinary bosons, $p=1$.

Circa 1970, for ordinary bosons Horn and Silver \cite{4}
constructed the analogous parameter-free statistical model, using
conserved-charge coherent states, and showed that it well
described inelastic $\pi ^{+}\pi ^{-}$ pair production from fixed
targets with laboratory kinetic-energies up to 27 GeV. The model
agreed with the universal trends of an experimental regularity
reported by Wang. Horn and Silver argued for a statistical
treatment of the gross features because (i) momentum conservation
should be a weak constraint since the emitted pions occupy a small
part of the available phase space, (ii) total isospin conservation
on the distribution of charged pions should also be weak since
neutral pions are summed over, and so (iii) charge conservation
remains as the important constraint.

A charged-paraboson pair in order $p=2$ consists of one A quantum
of charge `+1' and one B quantum of charge `-1'. The Hermitian
charge operator is defined by $Q=N_{a}-N_{b} $ where $N_{a,b}$ are
the parabose number operators. $Q$ does not commute with $a$ or $\
b$, and the paraboson pair operators $ab\neq ba$. Nevertheless,
since $\lbrack Q,ab]=0,\quad \lbrack Q,ba]=0,\quad \lbrack
ab,ba]=0 $, the $p=2$ conserved-charge coherent state can be
defined as a simultaneous eigenstate of $Q,ab,$ and $ba$:
\begin{equation}
Q|q,z,z^{\prime }>=q|q,z,z^{\prime }>,\quad ab|q,z,z^{\prime
}>=z|q,z,z^{\prime }>,\quad ba|q,z,z^{\prime }>=z^{\prime
}|q,z,z^{\prime }>
\end{equation}
Unlike in the $p=1$ case \cite{4} where only one parameter arises,
here two complex numbers $z$ and $z^{\prime } $ arise because $ab$
and $ba$ are fundamentally distinct operators. Consequently, in
the following multiplicity considerations, two non-negative
parameters occur which are the moduli of these two complex
numbers, $u\equiv |z|$ and $v\equiv |z^{\prime }|
$. These parameters can be interpreted as the intensity strengths of the ``$%
ab$" and ``$ba$" sources. The explicit expressions for the
conserved-charge coherent states $|q,z,z^{\prime }>$ are given in
Ref. [5]. For $q$ fixed, the percentage of events with $m$ such
pairs, $P_{m}(q)$, is the square of the moduli of the expansion
coefficients of $|q,z,z^{\prime }>$ in terms of the two-mode
parabose number Fock states.

The figure shows the pair-probability $P_{1}(1)$ for the production of two $%
A^{+}$ and one $B^{-}$. The formulas for the $p=1$ curve are for
the source-intensity-strength $x$ non-negative
\begin{eqnarray}
<n>=\frac{xI_{2}(2x)}{I_{1}(2x)},\quad <n^{2}>=<n>+\frac{x^{2}I_{3}(2x)}{%
I_{1}(2x)} \\
P_{1}^{(1)}(1) =\frac{x^{3}}{2I_{1}(2x)}
\end{eqnarray}
and for the $p=2$ ribbon are for the intensity strengths $u$, $v$
non-negative
\begin{eqnarray}
<n>=\frac{1}{2}(\frac{uI_{1}(u)}{I_{0}(u)}+\frac{vI_{2}(v)}{I_{1}(v)}) \\
<n^{2}>=<n>+\frac{1}{4}(\frac{u^{2}I_{2}(u)}{I_{0}(u)}+\frac{%
2uvI_{1}(u)I_{2}(v)}{I_{0}(u)I_{1}(v)}+\frac{v^{2}I_{3}(v)}{I_{1}(v)}) \\
P_{1}^{(2)}(1) =\frac{u^{2}v+v^{3}}{8I_{0}(u)I_{1}(v)}
\end{eqnarray}
where the $I_{\nu }$'s are modified Bessel functions.

In the figure, the solid line is the $p=1$ curve. This $p=1$ curve
is also the $\{0,v\}$ line on the $p=2 $ ribbon. Near the peak,
there is a ``fold" at the bottom edge of the ribbon. As shown, the
$p=2$ ribbon consists of open-circles for the non-folded $u \ge v$
region, and of solid-circles for the folded $u \le v$ region. The
upper edge of the ribbon is the $\{u,0\}$ set of points. Slightly
to the right of the peak, one can see from the solid-circles that
each line of dots travelling leftward down the page, bends under
(or ``over", whichever as the viewer prefers) the fold to reach
the $\{0,v\}$ line on the ribbon.

For the cases of $q=0$ and $q=1$, the $p=1$ curve always lies on
the $p=2$
two dimensional surface. For $q$ odd, as shown in the figure, the ribbon is $%
u \leftrightarrow v $ asymmetric and the ribbon is only partly
folded over. For $q$ even, there is a complete fold $u =v$ of the
ribbon because then the $u > v $ and $v > u $ surfaces are
identical.

The ``line of dots travelling leftward down the page" in the peak
region of the figure are a set of $\{ u, v \}$ values from a
unit-negative-slope diagonal in the $\{ u, v \}$ domain. In
independent-particle-emission models, the total energy in the
emitted particles is monotonically related
to the intensity strength of the source. This suggests that with the sum ${%
E^{total}}_A + {E^{total}}_B $ fixed, there will be a significant ${E^{total}%
}_A - {E^{total}}_B $ energy dependence in coherent paraboson
pair-production. ${E^{total}}_{A,B} $ are respectively the total
emitted $Q= 1$ and $Q=-1$ quanta's energies.

There is overall $A^{+} \leftrightarrow B^{-} $ symmetry ( $U(1)$
charge symmetry ): The ``Probability for $(m + |q|) $ $A^{+}$'s
and $(m)$ $B^{-}$'s " equals ``Probability for $(m) $ $A^{+}$'s
and $(m + |q| )$ $B^{-}$'s ",
because $P_{m}(|q|)=P_{m+|q|}(-|q|)$ and $%
<n_{b}^{M}>_{-|q|}=<n_{a}^{M}>_{|q|}$ for $M=$integer.
Nevertheless, for q odd, physical observables such as $P_{m}(q)$
are not symmetric under the $u \leftrightarrow v$ exchange of the
``$ab$" and ``$ba$" intensity strengths. See Eqs. (5-7).

This work was partially supported by U.S. Dept. of Energy Contract
No. DE-FG 02-86ER40291.

\vbox {\vspace{1cm}}

\section*{Figure Caption}

Figure 1: The pair-probability $P_{1}(1)$ for the production of
two $A^{+}$ parabosons and one $B^{-}$ paraboson versus the mean
number of charged-pairs $<n>$ and $<n^2>$.

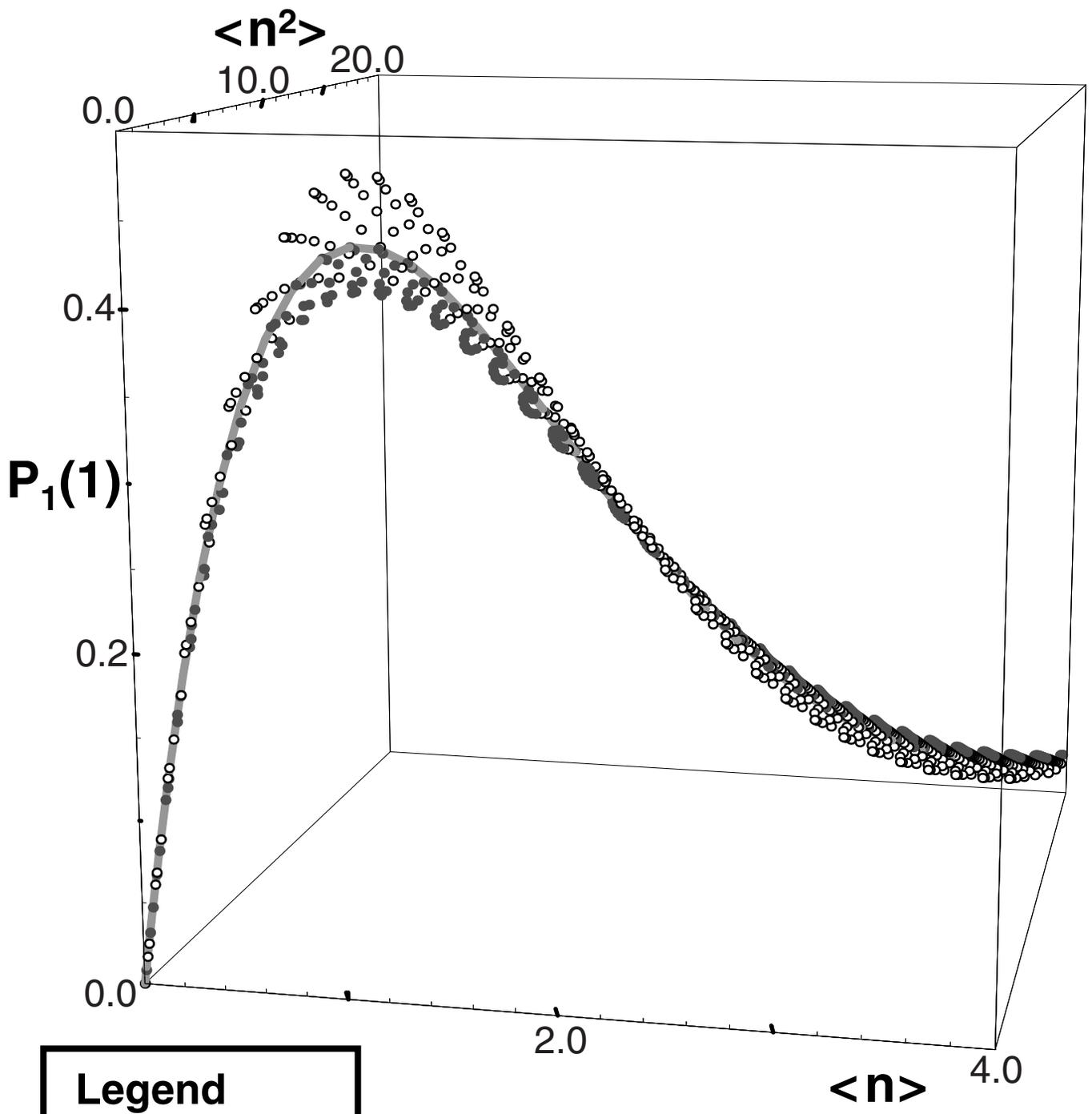
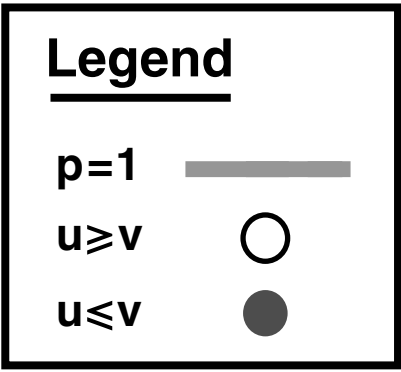
\end{document}